# Towards a Systematic View on Cybersecurity Ecology

Wojciech Mazurczyk[1], Szymon Drobniak[2] and Sean Moore[3]

[1] Warsaw University of Technology, Institute of Telecommunications, Poland, email: wmazurczyk@tele.pw.edu.pl
[2] Jagiellonian University, Institute of Environmental Sciences, Poland, email: szymek.drobniak@uj.edu.pl
[3] Centripetal Networks, USA, email: smoorephd@gmail.com

*Abstract* —Current network security systems are progressively showing their limitations. One credible estimate is that only about 45% of new threats are detected. Therefore it is vital to find a new direction that cybersecurity development should follow. We argue that the next generation of cybersecurity systems should seek inspiration in nature. This approach has been used before in the first generation of cybersecurity systems; however, since then cyber threats and environment have evolved significantly, and accordingly the first-generation systems have lost their effectiveness. A next generation of bio-inspired cybersecurity research is emerging, but progress is hindered by the lack of a framework for mapping biological security systems to their cyber analogies. In this paper, using terminology and concepts from biology, we describe a cybersecurity ecology and a framework that may be used to systematically research and develop bio-inspired cybersecurity.

## I. Introduction

It is estimated that current commercially available anti-virus products are able to detect only 45% of the new threats that Internet users face each day [1]. Moreover, the number and functionality of malicious software utilized by cybercriminals, as well as its sophistication and complexity, is constantly increasing. As a result, the average length of time between initial injection of a threat into the network and its discovery is increasing every year, and is now measured in months (according to Verizon's "2014 Data Breach Investigations Report"), if not years. Additionally, current defense systems are largely static and not sufficiently adaptable to cope with the attackers' changing tools and tactics. The inability to provide trusted secure services in contemporary communication networks could potentially have a tremendous socio-economic impact on both E2E and E2C global markets. Because currently available cyber defenses are progressively showing their limitations, it is imperative to find a new direction for cybersecurity research and development to follow.

We propose that the network security community should look into nature for new approaches to cybersecurity, both offensive and defensive. Current and future cybersecurity solutions should be designed, developed, and deployed in a way that will fully leverage the experience, learning, and knowledge from on-going biological evolution. Conversely, the community should also look to nature to anticipate how the threat may evolve, and respond accordingly.

The most notable pros and cons of the bio-inspired cybersecurity approach are detailed below.

First, nature has over 3.8 billion years of experience in developing solutions and adaptations to the challenges that organisms face living in extremely diverse environmental conditions. The estimated number of (largely undiscovered) species is tens of millions, and each of them possesses specific and unique traits facilitating survival and propagation of their own genes. The key process of living organisms that has led to the persistence of the most successful forms and behaviors is evolution. Evolution has developed optimal solutions for situations analogous to the threats faced by computer network systems.

Second, for ages people have sought inspiration from nature. Some relevant modern examples include biomimicry, which is the inspiration of such inventions as Velcro tape and "cat's eyes" (retroreflective road markings). Computer science has also taken a page out of nature's book by developing biologically inspired techniques like genetic algorithms, neural and sensor networks, etc. Although at first glance there may not appear to be a direct relationship between cybersecurity and the patterns present in nature, closer inspection reveals that the essence of most known Internet attacks and defence mechanisms has analogies in nature. For example the Kudzu vine is able to penetrate its ecosystem with an astounding speed of ca. 30cm/day. Within a short time it can choke all other vegetation, including trees and shrubs, by blocking access to the resources necessary for survival – light and nutrients. The essence is just like in DDoS (Distributed Denial of Service) attacks for communication networks where legitimate users are deprived of the resources that they are entitled to like access to the service, bandwidth, CPU time, etc. Similar analogies can be drawn for other offensive techniques as well as for security solutions, as observed and described in [2].

Another powerful analogy is the "arms race" (a form of a coevolution involving an aggressor developing its offensive mechanisms and a victim/host evolving countermeasures in the form of defensive barriers). "Arms race" is often observed between e.g. predators and prey in nature. Similar dynamics can be also found in interactions involving hosts and parasites, with the former constantly trying to invade host bodies and the latter constantly evolving countermeasures preventing the invasion. Both the abovementioned cases bear many resemblances with a "malware-security systems" scenario (or more generally "attackers-defenders") where there is a continual contention to develop offensive/defensive measures as fast as possible to at least temporarily dominate the other side. Thus, it is readily apparent that in both nature and cyber world, entities must evolve permanently and adapt to ever-changing environments. In biology this phenomenon – an organism's need to continually adapt and evolve to avoid extinction – is called the Red Queen hypothesis [19]. It was named after a character from Lewis Carroll's book "Through the Looking-Glass". In this book the Red Queen described her country as a place where "…it takes all the running you can do, to keep in the same place". Exactly the same process can be observed in cybersecurity and in biological systems where there is a constant need for adaptation of offensive/defensive techniques to maintain a certain level of adaptation permitting survival and reproduction/propagation.

Bio-inspired cybersecurity is not a new idea. The first generations of cybersecurity research were bio-inspired, e.g., the immune system inspired defense methods based on signature analysis, as well as methods for handling polymorphic threats (which are analogous to, e.g., different influenza strains). Since then, however, the threats have evolved to make these first-generation defenses less effective. In order to survive, cybersecurity must be evolved/adapted accordingly to counter the new threats. A next generation of bio-inspired cybersecurity research is now emerging; however, we find the knowledge and achievements to be scattered because the field lacks a framework. This paper aims at filling this gap by defining, based on the terminology and concepts known from biology, the ***cybersecurity ecology*** (and related terms). This cybersecurity ecology will enable a rigorous analysis of the existing relationships between entities in the ***cybersecurity ecosystem***. Such a systematic view of cybersecurity will allow the research community to analyze and compare biological organisms' interactions with those from the virtual world in order to identify differences, deficits and potentially new promising approaches to cybersecurity.

We need to be cautious, however, that the mappings from nature to the cyber world are not always "1-to-1", i.e., the analogies are not always perfect. Some of the reasons that exact mappings are not always possible include:
- Many mechanisms and relationships in nature are very complex and not yet understood sufficiently to correctly map them to the virtual world;
- In nature, individual organisms within a species are disposable, and death is a critical driver of evolutionary adaptation; but for many security-critical systems (e.g. military, utilities, and other critical infrastructure) any loss, compromise, or corruption is unacceptable;
- The main goal for any organism is to survive and reproduce, whereas our computers / networks have many different goals (specific tasks and functions).

Despite these imperfect mappings we strongly believe that there are still many important lessons from nature that can benefit and improve cybersecurity. Moreover, if we follow a Sapir-Whorf hypothesis [31], which states that language has a direct impact on thoughts, then finding analogies between cybersecurity and nature with its accompanying terminology, concepts and solutions can have a tremendous impact on the way we think about solving cybersecurity problems. New mechanisms and ideas may emerge. Therefore, the systematic view for bio-inspired cybersecurity that we are proposing should help to unveil new promising directions that could be pursued to discover and develop effective next-generation security solutions.

The rest of this paper is structured as follows. Section 2 summarizes the state-of-the-art in bio-inspired cybersecurity. In Section 3 the analogy between the biology-based ecosystem and the cyber-ecosystem, including potential interactions, is drawn. Section 4 describes some promising research directions for cybersecurity. Finally, the last section concludes our work.

## II. RELATED WORK

The existing literature includes many attempts to map biological concepts to cybersecurity. And, many of these attempts have successfully transitioned to cybersecurity technologies and systems in common use nowadays, including anti-virus, intrusion detection, threat behavior analysis, honeypots, counterattack, etc. [2]. As already mentioned in the previous section, current research on bio-inspired cybersecurity is fragmented and lacks a systematic approach. A primary cause is the diversity of aspects from nature that can be used as inspiration for cybersecurity research. Current research may be broadly segmented into two groups, depending on how an inspiration is drawn:
- when inspiration is drawn from a given organism's characteristic feature/defense mechanism (internal or external). Internal mechanisms include, for example, an immune system. External mechanisms include e.g. various camouflage and mimicry techniques;
- when inspiration is drawn from various inter-organism interactions – this includes, e.g., predator-prey associations.

*2.1 Bio-inspired cybersecurity inspired by an organism's characteristic feature/defense mechanism*

In order to effectively avoid detection/observation an organism can hide or conceal its presence by using camouflage or mimicry techniques that modify the organism's external appearance [17].

Camouflage embraces all solutions that utilize individual's physical shape, texture, coloration, illumination, etc. to make animals difficult to spot. This causes the information about their exact location to remain ambiguous. Examples of animals that can easily blend into the background include the chameleon (family *Chameleonidae*) which can shift its skin color to make it similar to ambient lighting and background coloration; stick and leaf insects (order *Phasmotodea*) that take the physical form of a wooden stick or a leaf; orchid mantis (*Hymenopus coronatus*) that resembles a tropical orchid which, although quite conspicuous, is difficult to detect against a background of developed flowers. Camouflage often occurs on levels other than visual recognition: e.g., many viruses code pathways and molecular signaling systems that mimic host cell transduction mechanisms – by doing so the virus can easily invade the cell and take control of the metabolism and immunological system of an individual [20]. In cyber space various information hiding techniques, e.g. steganography, can be utilized to provide means to hide the location of confidential data within an innocent-looking carrier or to otherwise enable covert communication across communication networks [18].

Patterns and/or colorations can be also used to confuse the predator, i.e., to make information about the prey hard to interpret. Such so-called "disruptive" camouflage is possible and can be seen in, e.g., a herd of zebras (*Equus quagga*) where it is difficult for an attacking lion to identify a single animal in a herd when they flee in panic. Patterns of contrasting stripes purportedly degrade an observer's ability to judge the speed and direction of moving prey, and they do so by exploiting specific mechanisms associated with the way brain processes visual information on movement [21]. An analogous idea is utilized by various moving target techniques/defense in cyberspace, which distribute the uncertainty between the attacker and the defender more fairly. For example, some first-generation solutions made periodic changes in a host's appearance from the network perspective, in order to mitigate the effectiveness of target reconnaissance [8]. Second-generation solutions include, e.g., an ant-based cyber defense which is a mobile resilient security system that removes attackers' ability to rely on prior experience, without requiring motion in the protected infrastructure [12].

Mimicry characterizes the cases in which an organism's attributes are obfuscated by adopting the characteristics of another living organism. In particular, this means that the prey can avoid attack by making the predator believe it is something else, e.g., a harmless species can mimic a dangerous one. The prey hides information about its own identity by impersonating something that it is not. For example, harmless milk snakes (*Lampropeltis sp.*) mimic venomous coral snakes (*Micrurus sp.*) to confuse predators which are less likely to launch an attack in expectation of a venomous harmful bite. Cybersecurity solutions that utilize the same idea include various traffic type obfuscation techniques, e.g., traffic morphing [16].

Organisms' internal systems may also inspire new cybersecurity approaches. There are many recent studies attempting to map features and functions of the human immune system to cyber space [3, 9, 10, 11]. Immune systems use a diversity of receptors to detect external antigens (alien proteins). These variations are not inherited but instead are generated via recombination in the process of V(D)J (somatic) recombination, which generates repertoires of receptors undergoing clonal selection and reinforcement – preparing them for effective action against antigens, with the lowest possible level of autoagression (e.g. reaction against an organism's own proteins) [22]. The resultant Artificial Immune Systems (AIS) are designed to mimic certain properties of the natural immune system. In cybersecurity their main application is anomaly and misbehavior detection. AIS typically rely on one of four major paradigms: *(i)* negative selection algorithm [3]; *(ii)* clonal selection algorithm [9]; *(iii)* dendritic cell algorithm [10] or *(iv)* idiotypic networks models algorithms [11]. The first generation AIS (*i* and *ii*) utilized only simple models of human immune systems, so the resulting performance was not comparable with its human counterpart. Recent AIS (*iii* and *iv*) are more rigorous and better correspond to natural immune systems.

*2.2 Bio-inspired cybersecurity inspired by organisms' interactions*

In nature, there are many interactions between organisms that potentially may serve as inspirations for cybersecurity.

For example, several studies focus on various aspects of predator-prey associations. In [13] the authors make the predator-prey analogy for the Internet and investigate how different levels of species diversification can serve as a defensive measure. They considered each type of a vulnerable device as a heterogeneous species and investigated what level of species diversification is necessary to prevent a malicious attack from causing a failure to the entire network. Subsequently, in [5] it was discovered that the cost to the predator in seeking its prey drastically impacts the predation process. In particular it has been observed that even fairly simple strategies for raising the cost of predation can result in significant reduction in outbreak size. Other studies utilize biological models of epidemic spreading (a special case of antagonistic interaction between the pathogen and the victim) to predict or analyze malware outbreaks [14], [15].

Finally, the relationships and interactions between existing malware (so called malware ecology) have been investigated in [6]. Numbers of interactions, both accidental and intentional, between different types of malware were analyzed and the main conclusion was to seek ecologically-inspired defense techniques, because many ideas from ecology can be directly applied to all aspects of malware defense.

From the studies presented above we can conclude that bio-inspired cybersecurity is a wide, diverse, emerging, and evolving research field. However, from the research perspective, we see many "loose ends" that need to be tied by using a more systematic approach, which we next propose.

III. CYBERSECURITY ECOLOGY

In this section, first we systematically review the key terms from biology related to ecology. Then by borrowing and adjusting the original biology-based definitions, we will describe the most important components of cyber-ecosystem and then of cybersecurity ecology.

*3.1 Cyber-ecosystem*

In biology the term *ecology* is defined as the field of life sciences analyzing and studying interactions among organisms and/or their environment. This means that it deals with the structure and functioning of *ecosystems*. An ecosystem is defined as a community of living organisms (biotic components) together with the nonliving (abiotic) components of their environment that interact as a system. Apart from the biotic and abiotic components, interconnected by various interactions, the ecosystem is fueled by energy, usually in the form of electromagnetic radiation (if production in an ecosystem is sun-driven, i.e.

accomplished by green plants) and chemical energy (if an ecosystem relies on chemosynthetic bacteria). Both biotic and abiotic factors can influence an organism. For example, climate change or an atypically large number of predators can negatively impact some species [23].

In every ecosystem the energy flow is crucial as each ecosystem is energy-based and is capable of transforming, accumulating, and circulating energy. In nature the flow of energy is encapsulated in a food chain, and a concept of trophic levels is utilized to illustrate the position that an organism occupies in a food chain (Fig. 1, left). Depending on how energy is obtained, two groups of organisms can be distinguished: producers (that are able to manufacture their own food using inorganic components and chemical/radiation energy) and consumers (that feed on producers and/or other consumers) [24].

Ecology can be viewed as one of the approaches to study complex and dynamic systems. Thus, if we are able to understand how ecosystems and related concepts map to the cybersecurity field then the usefulness of various ecological methodologies can be evaluated. If such mappings are successful then application of many mathematical ecological systems models to cyber systems can be investigated.

Based on the above terms and definitions from ecology, we want to systematically recreate an analogous taxonomy for the cyber world.

Let us define *cyber-ecosystem* as a community of *cyber-organisms* i.e. non-human actors e.g. applications, processes, programs, defensive and offensive systems (analogues to the biotic components) that interact between themselves and with the environment (abiotic components). Let us also assume that the environment in which biotic components reside and interact is a communication network, e.g. the Internet, and it constitutes a nonliving (abiotic) component with its hardware, links and interconnections.

In the cyber-ecosystem (the same as in nature) both biotic and abiotic factors can impact a cyber-organism. For example, malicious software can be utilized to compromise a user's device defenses and steal his/her confidential data. On the other hand a failure of the link/networking device or network congestion influence a cyber-organism's ability to communicate and exchange information.

In such a defined cyber-ecosystem we are particularly interested in the network of interactions among cyber-organisms, and between cyber-organisms and their environment.

As mentioned above, in nature the key resource is energy. In communication networks, the analogous key resource is different kinds of *information,* including user personal or user-generated data, but also information about his/her behavior. In such a cyber-ecosystem, information can be transformed, accumulated, and/or circulated (similar to energy in ecosystems).

To have more clear analogies between ecosystems and cyber-ecosystems the role of the humans in the present context is constrained to these roles:
- *Producers* which possess and generate information that forms a desirable resource for the consumers (e.g. the tools that attackers or digital marketing companies use to obtain desired information).
- *Components* of the offensive/defensive solutions. For example, a bot herder typically issues command to the bot that he controls so he is an inevitable "part" of the botnet. Another example is an ID/PS (Intrusion Detection/Prevention System) which is configured and monitored by a security specialist.
- A part of *"evolutionary force"*. Humans influence cyber-organisms by changing their code, functionalities and applications. In this way an evolution is achieved. Typically, attackers try to outwit the defenders by developing malicious software that will be capable of overcoming existing defense mechanisms/systems. Conversely, defenders develop their defenses to be "immune" to the existing threats. Thus, both sides are taking part in a cyber "arms race".

Considering the above, it is possible also for the cyber world to characterize certain "cyber food chains" and/or cyber-trophic levels (Fig. 1, right). Consumers can become cyber-predator (attacker) or cyber-prey (defender) depending on the location in the cyber food chain. Producers always take the role of cyber-prey.

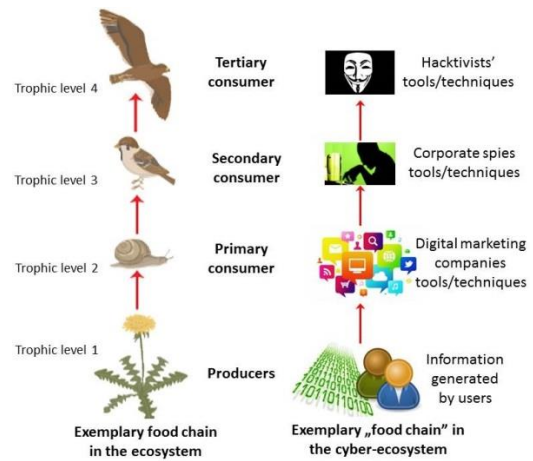

Fig. 1 Food chains and trophic levels in an exemplary ecosystem (left) and a cyber-ecosystem (right).

*3.2 Cyber-ecology and its subtypes*
By means of a simple analogy we can define the following terms that rigorously describe the toolbox of cybersecurity ecology:
- *Cyber Ecology* (CE) as a field that analyzes and studies interaction among cyber-organisms and/or their environment.
- *Cybersecurity Ecology* (CSE) analyzes and studies interactions among cyber-organisms and between cyber-organisms and their environment that influence their security. CSE is a subfield of CE.
- *Attacker–Defender Ecology* (ADE) describes interactions between cyber-organisms which take roles of attackers and defenders in the specific cyber-ecosystem (e.g. in the

Internet). As noted before such relationship can be regarded not only as predation but also as parasitism. It is also worth noting that such interactions reside in different locations of the cyber food chain and depend on the trophic level (Fig. 1). ADE is a part of CSE.

- *Attackers Ecology* (AE) illustrates interactions between attackers (cyber-organisms) in a given cyber-ecosystem. The possible interactions encompass both antagonistic and non-antagonistic ones and depend on the context. Attackers can predate or parasite on each other, but the relationship can be of a symbiotic or a cooperative nature. AE is a part of CSE.
- *Defenders Ecology* (DE) provides insights into potential interactions between the defenders (cyber-organisms), and it incorporates mostly non-antagonistic ones. It includes both external defense mechanisms (interactions of malware and defense systems resulting in defense) and internal properties (analogous to animal immune systems). DE is a part of CSE.

The abovementioned terms e.g. AE can be further divided into e.g. malware ecology, botnet ecology, etc. The relationships between the terms defined in this and in previous sections are illustrated in Fig. 2.

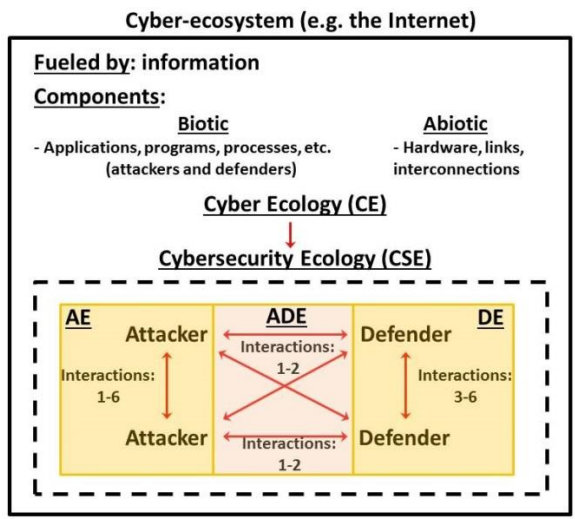

Fig. 2 Main components and interactions in a cyber-ecosystem (Interactions: 1-predation, 2-parasitism, 3-symbiosis, 4-cooperation, 5-sexual interactions, 6-competition).

*3.3 Cyber-ecosystem interactions*
The structure and stability of an ecosystem in nature is determined by the set of *interactions* that interconnect different entities. Interactions can be roughly classified into antagonistic interactions (between species; mainly predation and parasitism), non-antagonistic interactions (between and within species; cooperation, symbiosis) and sexual selection-driven interactions (within species). In all three classes, interacting entities coevolve, responding reciprocally to their current states in a positive/negative feedback loop mechanism (also known as the arms-race dynamics for antagonistic interactions) [23]. The interactions can be defined as follows:

- *predation*: a way of obtaining resources by killing/eating bodies of other organisms; results in the death of the prey; predation involves complex cycles of prey and predator abundances described by mathematical models such as the Lotka-Volterra equations system ([23], [25]), which can be utilized to design the most optimal strategies of defense or offense, depending on which side of the predation-prey system the focal cyber-organism currently is. In communication networks ransomware can be treated as a predator as it is "killing" the host by encrypting vital information it stores and unless the ransom is paid this resource is "destroyed/lost" i.e. user's data cannot be retrieved;
- *parasitism*: interaction involving obtaining resources by eating other entities but not killing them [23], [25]; it gave rise to a fruitful field of epidemiological parasitology, with mathematical models and defense systems that could be directly implemented in the context of cyber-epidemics. As already mentioned the current trend, especially for sophisticated malware such as Advanced Persistent Threats (APTs), is more similar to a parasite-host scenario than a predation-prey one. It means that it is more likely that the malicious software will be active on an infected host for a long time and obtaining its resources in a transparent manner;
- *symbiosis*: positive interaction involving obligatory interaction of two or more entities, necessary for all parties for survival and successful propagation. In cybersecurity this could include analysis of both attackers and defenders symbiosis. For example, for malware infection scenario it is common that the first infection is initially performed by exploiting some vulnerability on the host machine and this allows later for the second part of malware to be downloaded and executed in order to perform malicious actions for the cybercriminal;
- *cooperation*: facultative interaction of an individual within one species or members of different species, increasing the fitness and survival of other individuals (the acceptors of cooperation) often at the cost of the focal individual (the giver of cooperative behavior) [23], [26], [27]; in communication networks cooperation should be recognized not only as a way of reinforcing defense mechanisms but also as a potential threat (a deceiver malware might exploit cooperating inclination of the system, wreaking havoc in its structures). A recent real-world example is the sharing of cyber threat indicators as prescribed in the US Cybersecurity Information Sharing Act of 2015;
- *sexual interactions*: occur exclusively within species and are channeled toward combining, in the most desired and effective way, the genes of females and males so that they maximize the fitness of offspring [28]; from the point of view of cyber-ecosystems the models of sexual selection based on compatible genes [29] are particularly interesting as they may serve as mechanisms for producing dynamic sets of the most optimal combinations of entities and their mutations that provide maximum protection against evolving malware. Moreover, using knowledge of how sexual selection works, it may be interesting to study how to become the most "unattractive" victim to the potential attacker.

- *competition*: this relationship is symmetrical and involves both organisms competing for the same pool of resources. Inherently the relationship between organisms can be broken without any harm to neither of the sides – as both influences are negative their cessation benefits both competitors. In communication network environment this interaction can occur e.g. between two types of malware trying to infect the same host – when one of them succeeds it tries to "secure" the host by patching the exploit used by the other type of malicious software. Competition can also occur between defenders when few similar defense systems (e.g. anti-virus software) are run together and they impact each other in a negative way.

A point of view of cyber-ecology may be to treat these interactions as purely mechanistic descriptions of cyber-systems – without looking at the consequences of interactions themselves and on the dynamics they describe. However, growing evidence suggests that the interactions not only influence the fitness and performance of entities but also significantly modify their physiology/performance in the interaction, altering the outcome of competition/synergy [30]. Such elastic responses of interacting entities to the interaction itself may have a significant role in cyber-ecosystems, as they may serve to design more efficient ways of controlling cyber-ecosystems and reacting to unknown, emerging threats.

As indicated in Section II, existing work focuses mainly on predator-prey association. However, an interesting observation is that the relationship between current malware and host is in essence closer to parasitism than to predation. This means that the goal of the current malware is to live off the infected host (and the longer it remains undetected, the better) but not to immediately cause significant harm or permanent damage.

In the following section we will review the most important natural-enemy ecology models including parasitism models, and we will assess how this knowledge can be used for cybersecurity purposes.

## IV. Natural Enemy Ecology in Nature – unifying antagonistic interactions

The field of antagonistic interactions in ecological studies has so far been dominated by a very sharp distinction between predator-prey interactions and parasite-host interactions. As pointed out recently such interactions are, however, much closer to each other, and together with a third class (competition) form a unified group of antagonistic interactions involving the aggressor, the victim and resources that are/may be available to one or both entities [32]. This has led to the emergence of a new field-of-study in ecology, which is broadly termed "natural enemy ecology", and encompasses all interactions involving detrimental effects of one organism on another, be it a direct or indirect (e.g. via shared resources) effect. In this section we discuss consequences of such a categorization and review the most prominent models of antagonistic interactions, while pinpointing their weaknesses [32].

*4.1. Similarities between parasitic and predatory interactions*

The strong distinction between parasitic and predatory relationships results mostly from an old methodology of categorizing nature [33]. In fact, all kinds of antagonistic ecological interactions (predation, parasitism and competition) share a common suite of components, which differ only in the strength/presence/direct character of the specific connections. All interactions involve conventionally at least two organisms (aggressor and victim, or two competitors in the competition model) that influence each other positively and/or negatively, and use each others' resources [32].

**Competition**: the least antagonistic of all interactions; the roles of the interacting organisms are indistinguishable and both exert mutually negative influence on the other. The relationship is symmetrical and involves both organisms competing for the same pool of resources. Inherently the relationship between organisms can be broken without any harm done to neither of the sides: as both influences are negative their cessation benefits both competitors [32].

**Predation**: occurs when the aggressor kills the victim directly and feeds on its tissue – therefore it is inherently asymmetrical; predation involves very short time-scales, much shorter than timescales necessary for the evolution of low-level (molecular, immunological) defense mechanisms and, thus, prey evolves defenses in such system mostly at the higher, organismal (e.g. morphology and behavior) level [34]. Instead of immunological mechanisms prey benefits more by evolving learning-like mechanisms that are much more flexible on one hand and can evolve within long generation times on the other hand. Because predators consume their victims, they are regarded as residing on a different, higher trophic level than prey [32].

**Parasitism**: in this form of interaction the aggressor feeds on the victim but does not kill it. Predatory interactions are inherently fatal whereas parasitic interactions have led to the phenomenon of intermediate virulence, which maximizes parasite transmission to other hosts. The relationship between parasites and hosts is much more intimate and occurs at time-scales and generation times that allow the evolution of complex genetic (e.g. bacterial Crispr-Cas [35]) and immunological (e.g. vertebrate acquired immunity, invertebrate Toll receptors) defense mechanisms in victims/hosts.

It is clear that all three relationships are slightly different and involve different levels of inter-organismal contact. However they all draw from the same population processes related to population growth and decline. Moreover, sometimes parasitism and predation are hard to delineate. For example, caterpillars feeding on plants could be regarded as predators, but they do not kill their victims and dwell on the surface of victim, as ectoparasites. Mosquitos feed on the tissues of their victims (like parasites) but apart from this they display many properties of predators (longer generation time, short interaction timescale, high turnover rate of attacked victims). Recent literature has also pointed out that although seemingly different, parasitic and predatory interactions may give rise to similar ecological patterns. Some prominent examples include:

- *The evolution of inducible defenses and attack anticipation* [36]: predation is often associated with behaviors and traits that are active and use resources only in the presence of predators – similar mechanisms may be present in the parasite-host systems where organismal systems (e.g. immunological) may optimize their activity window to match the activity window of aggressors,
- *Enemy-mediated facilitation* [37]: in the presence of more than one aggressor, host/prey communities may evolve mechanisms that make use of prey-specific resistance to aggressors and indirect ecological effects that result from variation in prey/host susceptibility to aggressors,
- *Managing the threshold of transmission*: in parasite-host systems there are specific host densities below which parasites are unable to effectively spread and persist; a similar concept might be applied to the predator-prey systems, where by managing the densities of particular predators ("superpredators" that affect prey densities the most) the population may be maintained at a desired level of prey density, avoiding extinction due to random fluctuations in predation pressures [32].

*4.2. Models of antagonistic interactions*
The ecological literature has developed a number of mathematical descriptions of the predator-prey or parasite-host interactions and not surprisingly, and in line with the abovementioned unifying considerations, all these models can be adjusted for the description of both predation and parasitism interactions. The most prominent and the oldest model is the Lotka-Volterra (L-V) model [33] that binds together aggressor and victim densities and models changes in these densities according to an assumed predation/parasitism rate. The model is defined using a system of two differential equations:

$$\frac{dx}{dt} = rx - ayx$$

$$\frac{dy}{dt} = -r'y + a'xy$$

where *x* and *y* denote prey and predator densities, *r* and *r'* describe population growth/decline of prey/predator populations, whereas *a/a'* quantify the rate of encounters between prey and predators. The solution of this system describes the oscillatory behavior of prey and predator densities. The L-V model was quickly considered simplistic (e.g. the assumption of constant encounter rates *a/a'* was considered as biologically unrealistic) and a number of other models have been developed. However, ecologists agree that all available models are just special cases of the L-V model, which in turn still remains the most important model for antagonistic interactions among organisms [33].

The models that followed the L-V system focused mostly on making some of its assumptions more realistic. For example, the Nicholson-Bailey model expanded on the results from the L-V system and generalized them to discrete generations of prey and predators (the L-V system was developed under the assumption of continuous overlapping generations). More advanced models, e.g. the Holling model [38], the Ivlev model [39], and the Watt model [40] remained in the reality set by the Lotka-Volterra model, changing and adjusting only the encounter function (i.e. the function that binds prey and predator densities together with time, providing the dynamics of the encounter rates between interacting individuals).

A proper integration of the existing models into the field of cybersecurity will likely involve a revision of the assumptions of different models of antagonistic interactions and relating them to the specific features of communication networks. Specific comparisons are necessary to elucidate the shared features and assumptions at the interface of biological and cyber systems – such comparative analysis can then identify models that are the most accurate in describing cyber reality with respect to the antagonistic interactions.

*4.3 Antagonistic (parasitic) mimicry: Batesian mimicry*
Even without clear exploitation of material resources of the hosts, parasitism can be present if information content/reliability is being exploited by one organism at the expense of the costs born by the other organism [41]. One well-documented example of such behavior is parasitic mimicry, which is relatively inexpensive to the mimicking organism as it is not associated with weapons/toxins this organism is pretending to have [42]. A well-known example is the *Chrysotoxum festivum* hoverfly that resembles toxic and stinging insects from the Hymenoptera group. By expressing warning colors the hoverfly avoids being attacked and eaten, and on the other hand it does not have to invest resources in actually having a sting.

Parasitic (Batesian) mimicry, due to its inexpensive nature, could readily be used in security applications in cyber systems. The mimic could be the security algorithm that could adopt some features of the actual hostile software to approach it and infiltrate without being detected [41]. Most existing models of Batesian mimicry operate on the balance between costs of being detected and the benefits of expressing certain masking phenotypes. Such models could be used to derive parameter ranges that ensure full masking in the cyber-ecosystem at the expense of the lowest possible resource allocation.

*4.4. Non-antagonistic interactions*

Non-antagonistic interactions are more difficult to classify and organize, mostly because they combine intra- and inter-species processes. There exists no single model of synergistic interactions similar to the seminal Lotka-Volterra model; however, we have several ways of expressing the dynamics of such interactions mathematically. Non-antagonistic interactions that play major roles in development of cybersecurity solutions encompass all of the above sexual selection/mate choice processes, and symbiotic interactions. Both have the potential to substantially inform efforts to develop effective cybersecurity strategies; both also remain largely unstudied on a large, inter-species comparative level and thus are attractive targets of comparative biological research.

*4.5 Symbiotic interactions*

Symbiosis is thought to underlie all life on Earth as, according to the endosymbiosis hypothesis, all eukaryotic cells are descendants of several prokaryotic organisms that merged together as symbionts, which gave rise to currently observed organelles such us chloroplasts and mitochondria [43]. Currently the most commonly known and well-studied examples of such interactions may serve as good models to derive mathematical parameters that can be used in developing cybersecurity solutions. From an evolutionary perspective, the symbiotic interactions can be readily modelled using the same mathematical reasoning as the one used in the Lotka-Volterra system, by modifying parameters of the equations so that interacting units benefit each other instead of harming [44].

From the point of view of cybersecurity applications, symbiotic interactions may potentially play roles in two scenarios. For one, symbionts in a cyber-ecosystem could be used to strengthen the protective/immunizing effects of applied techniques. Multiple symbiotic entities could enforce each others' defensive strategies and achieve fuller protection of the whole system. On the other hand, symbiotic interactions are intricately associated with other close interactions. In fact, the Lotka-Volterra-like model of symbiotic interactions [44] predicts that they can easily turn into parasitic interactions if conditions shift in the environment of symbionts (e.g. if available resources become more asymmetrically exploited by one of the symbionts). Thus, such models are also able to provide a testing space where a range of parameters that maintain the beneficial symbiotic interactions could be tested. In fact, such models can also be used to derive alternative scenarios of fighting cyber parasites – if it is possible to "mutate" them and modify their responsiveness to the environment – changing a parasitic interaction into a symbiotic one with an artificially introduced additional organism [45].

A special case of synergistic interactions occurs in cooperating organisms when individuals bear costs (often the highest fitness costs, i.e. by postponing/entirely abandoning reproduction) and benefit other individuals by helping them (usually in the form of raising their offspring) [27]. The dynamics of such interactions is best known in the altruistic forms of cooperation, where it is predicted and described by the Hamilton inequality [46] that binds costs of the donor, benefit of the receiver, and their coefficient of relatedness that defines how costs and benefits are balanced on both sides of the interaction [46], [47]. In the context of this project, however, it is of a marginal importance – much more important kinds of cooperating interactions will be those encountered between non-related individuals. Such non-kin cooperation can easily be incorporated in our system (as reciprocal sharing of costs and achieved benefits), however this field of ecology is still strongly underrepresented and no quantitative models exist that could be used and developed in the context of the proposed project.

*4.6 Sexual selection*

From the point of view of cybersecurity, sexual selection may be the most difficult but also the most potent interaction that could be exploited [48]. The biggest difficulty comes from the fact that sexual selection operates through choice of the most suitable mates and thus would require creating and maintaining a population of sexually reproducing entities that would use cycles of selection in order to evolve new, more effective ways of fighting enemy software [28]. It is an important question how such selection would operate and currently evolutionary biology describes two major classes of sexual selection mechanisms.

The first one, called "the good genes hypothesis" poses that selective individuals (in nature usually females) choose certain partners (usually males) because they provide them with "good genes" that increase offspring viability and fitness [49]. Such indirect genetic benefits have been demonstrated in many animal studies and are a well-documented, although still weakly understood phenomenon [28], [29].

The second class of sexual selection drivers falls into the "Fisherian runaway" category, where the preference of one sex (females) evolves as a self-perpetuating mechanism that exploits certain male traits and is fueled by a positive feedback loop generated by the strong genetic correlations between female preference and male display traits [48], [49]. This second form of sexual selection has also been suggested to occur in nature – however it is much more difficult to find its place in the cybersecurity reality as this form of sexual selection is not directly associated with any fitness benefits to females (apart from choosing males that can actually afford to have exaggerated and overgrown traits).

Both models of sexual selection are governed by one common mathematical model [50] that integrates female preference (*P*), male display (*D*) and residual fitness effects (*F*). If we denote variance and covariance of specific traits as *V* and *C* (e.g. $V(P)$ – variance in preference; $C(PD)$ – covariance between display and preference), $b\_s$ and $b\_n$ as respective selection gradients resulting from sexual (s) and natural (n) selection, the joint dynamics of these traits may be described as:

$$\Delta \begin{pmatrix} \bar{D} \\ \bar{P} \\ \bar{F} \end{pmatrix} = \begin{pmatrix} V(D) & C(PD) & C(FD) \\ . & V(P) & C(FP) \\ . & . & V(F) \end{pmatrix} \times \begin{pmatrix} b\_n(D) \\ b\_n(P) \\ b\_n(F) \end{pmatrix} + \begin{pmatrix} b\_s(D) \\ b\_s(P) \\ b\_s(F) \end{pmatrix} + \begin{pmatrix} u(D) \\ u(P) \\ u(F) \end{pmatrix},$$

where *u* denotes respective changes in phenotypes' values due to mutation. Different combinations of parameters of this model yield different modes of sexual selection, and exploration of these values within the ranges that are realistic to cyber systems will help uncover types of interactions that would be the most efficient in cybersecurity applications.

## V. POTENTIAL BIO-INSPIRED RESEARCH DIRECTIONS FOR CYBERSECURITY

After defining key terms related to cybersecurity ecology, and describing most important models that characterize interactions between organisms in nature, the next step is to develop a "procedure" that will result in the potential new research directions. The steps of such a procedure related to interactions are illustrated in Fig. 3.

First, it is important to map existing offensive/defensive measures as well as interactions in both types of ecosystems. From the biology perspective this includes performing rigorous meta-analyses describing comparatively and phylogenetically the diversity of defense/offense mechanisms present in nature and their complexity (e.g. their costs, the most optimal uses, their diversity at various level of life organization).

In the next step, the missing components in the virtual world that could be potentially ported from nature should be identified. All of the most promising candidates that do not have counterparts in cyber space will form a list of most suitable bio-inspirations.

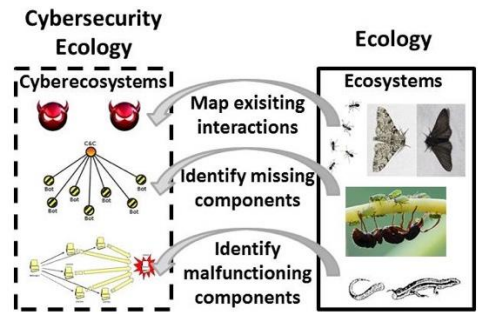

Fig. 3 Comparing interactions and components between ecology and cybersecurity ecology.

In the last step, it is also possible to identify security-related components that exist in cybersecurity but that are not sufficiently effective. Then, insights from mechanisms and relationships that exist in nature could provide important feedback on how these security techniques could be improved.

To summarize, we believe that currently the most promising research directions include:
- Drawing further inspirations from the particular organism's characteristic feature/defense mechanism. For example, such features like *aposematism* (warning signal that is associated with the unprofitability of a prey item to potential predators) or *autotomy* (where an animal sheds or discards one or more of its own body parts to elude or distract the predator) could readily become an inspiration for future cybersecurity solutions.
- Careful investigation and applying knowledge from the mentioned nature-based interactions. As already observed the malware-host scenario is more similar to *parasite-host* than to predator-prey association. Therefore, more research attention should be turned to the models and achievements of biology in this field. This could provide many new, interesting insights. Another research direction that we believe has not been sufficiently explored is *sexual interactions* where e.g. the methods to become an attractive/unattractive target could be analyzed.
- Comparative analysis of the features of parasitic and predatory systems that expose their common underlying mechanisms leading to their description within the natural enemy framework. Such common properties of these antagonistically interacting systems may be the most effective points (in a way identified by long evolutionary history of such systems) where new approaches to cybersecurity can be developed. The most promising avenues in this group of issues include *(i)* induced/anticipatory mechanisms that lower the costs of maintaining active defense mechanisms; *(ii)* enemy-driven facilitation – which, by exploiting multiple enemies, may lead to the establishment of reinforcement mechanisms that increase the effectiveness of enemy elimination; *(iii)* transmission threshold management which can provide tools to minimize the effort in eliminating threats, while maximizing the achieved security gain.

## VI. CONCLUSIONS

This paper presents a systematic ecology-based approach to cybersecurity. Based on the observation of the significant fragmentation of achievements and knowledge in the field of bio-inspired cybersecurity, we propose a cyber-ecosystem, cybersecurity ecology, and related terminology that may be used to study offensive/defensive mechanisms and interactions among cyber-organisms and/or between cyber-organisms and their environment. In our opinion this helps to identify new potential future research directions for bio-inspired cybersecurity.